# Penerapan E-Service Berbasis Android pada Divisi Pelayanan Perbaikan Komputer CV Ria Kencana Ungu (RKU)

Febri Valentina[1], Leon Andretti Abdillah[2], Nurul Adha Oktarini Saputri[3]

[1,2] Information Systems, Bina Darma University
[3] Informatics Engineering, Bina Darma University
Jalan Ahmad Yani No.3, Plaju, Palembang
[1] febri12142048@gmail.com, [2] leon.abdillah@yahoo.com

**Abstract.** Archival information systems in government agency is one of the most used applications for daily acitivities. One feature in application management information document is searching. This feature serves to search for documents from a collection of available information based on keywords entered by the user. But some researches on a search engine (searching) concluded that the average user error in the search is quite high due to several factors. Therefore, we need a development on this feature as search suggestion. This study discusses the application of the method of approximate string matching algorithm using levenshtein distance. Levensthein distance algorithm is capable of calculating the minimum distance conversion of a string into another string to the optimum. Archiving information system using Levensthein Algorithm String is an application that will be built to address these problems, this application will help, especially in the administration to enter or save a document, locate and make a report that will be seen by government agencies.

**Kyewords:** e-Service, Android, RKU.

## 1  Pendahuluan

Teknologi Informasi dan Komunikasi (TIK) banyak memberikan dampak dalam peningkatan nilai kompetitif perusahaan. Sementara teknologi informasi (TI) telah merambah ke *smartphon*e [1], salah satunya yang berbasis Android [2, 3]. Dengan menggunakan perangkat *mobile*, informasi dapat diperoleh dengan mudah dalam waktu singkat [4]. Aplikasia mobile telah banyak dikembangkan, seperti : 1) Membaca Iqro' [5], 2) Reservasi tiket [1], 3) m-Dictionary [2], 4) Lokasi pemukinan berbasis GIS [3], 5) serta Lokasi wisata alam [4]. Pada penelitian kali ini, penulis mengimplementasikan aplikasi *mobile* untuk *e-service*. *E-service* merupakan suatu aplikasi yang memanfaatkan TIK yang memiliki unsur penyedia layanan, penerima layanan dan pendukung pelanggan [6]. *E-Services* memiliki dua karakteristik utama : 1) Layanan ini dapat diakses dengan jaringan elektronik, dan 2) Layanan dikonsumsi oleh orang melalui internet [7]. Pada artikel kali ini, penulis akan menerapkan konsep *e-service* dengan perangkat cerdas, *smartphone Android*.

CV Ria Kencana Ungu (RKU) merupakan perusahaan yang bergerak dibidang penjualan dan perbaikan komputer. Selain menjual komputer CV RKU juga menjual *accessories* komputer, *mouse*, *keyboard*, *flashdisk*, *printer*, *catridge* dan lain-lain. Divisi pada CV RKU terbagi menjadi 4 (empat), yaitu : 1) divisi *marketing*, 2) divisi *printer*, 3) divisi *software*, dan 4) divisi *hardware*.

Tugas pada divisi *marketing* yaitu memasarkan produk kepada konsumen untuk menghubungkan antara produsen dengan konsumen sebagai pemakai produk, sedangkan tugas pada divisi *printer* yaitu memperbaiki *printer* dan memasarkan produk *printer* kepada konsumen. Untuk divisi *software* memiliki tugas untuk memperbaiki *software* yang rusak dan memberikan solusi yang tepat pada kerusakan *software* tersebut. Sedangkan divisi *hardware* yaitu memiliki tugas untuk melakukan perbaikan pada kerusakan perangkat keras berdasarkan kerusakan dan permintaan dari pihak terkait.

Proses bisnis yang dilakukan untuk pelayanan *service*, konsumen datang langsung ke CV RKU dengan membawa perangkat komputer yang memiliki masalah, kemudian divisi *service* akan mengecek apakah perangkat yang bermasalah itu dapat dikerjakan langsung atau butuh waktu yang lama, jika proses perbaikan perangkat tersebut memerlukan waktu yang lama, maka konsumen akan diberikan bukti perbaikan dan akan dikonfirmasi melalui media telepon ke konsumen jika perangkat yang bermasalah sudah diperbaiki. Permasalahan yang sering muncul yaitu konsumen kesulitan memperoleh perkembangan perangkat yang sedang di *service* dan juga kebiasaan konsumen yang sering mendesak pihak RKU mengenai perbaikan perangkat konsumen, padahal perangkat yang digunakan sedang dalam proses perbaikan. Dengan pemanfaat *e-service* ini diharapkan dapat memudahkan dalam memperoleh informasi *service* perangkat pelanggan.

Dengan permasalahan ini peneliti akan membuat sebuah aplikasi pelayanan yang dapat diakses dengan menggunakan perangkat bergerak (*mobile*) seperti *handphone*. Pelayanan ini dapat diterapkan dikarenakan perkembangan perangkat lunak mengalami perkembangan yang sangat signifikan dalam beberapa tahun terakhir. Dengan munculnya *platform mobile device* berbasis *android* yang saat ini mengalami peningkatan yang cukup pesat terhadap permintaan komunikasi data, dilihat dari segi layanan, kehandalan sistem, maupun laju transmisinya.

## 2 Metodologi Penelitian

### 2.1 Metode Pengumpulan Data

Data yang dibutuhkan dalam penelitian ini didapat dengan melibatkan sejumlah cara, yatiu : 1) observasi atau pengamatan, 2) wawancara atau *interview*, dan 3) studi kepustakaan.

### 2.2 Metode Pengembangan Sistem

Menurut Pressman [8], *prototyping* merupakan proses pembuatan *software* yang yang bersifat berulang dan dengan perencanaan yang cepat yang dimana terdapat umpan balik yang memungkinkan terjadinya perulangan dan perbaikan *software* sampai dengan *software* tersebut memenuhi kebutuhan dari si pengguna, dimana mengijinkan pengguna memiliki suatu gambaran awal/dasar tentang program serta melakukan pengujian awal yang didasarkan pada konsep model kerja. Metode pengembangan *prototype* dapat dilihat pada gambar 1.

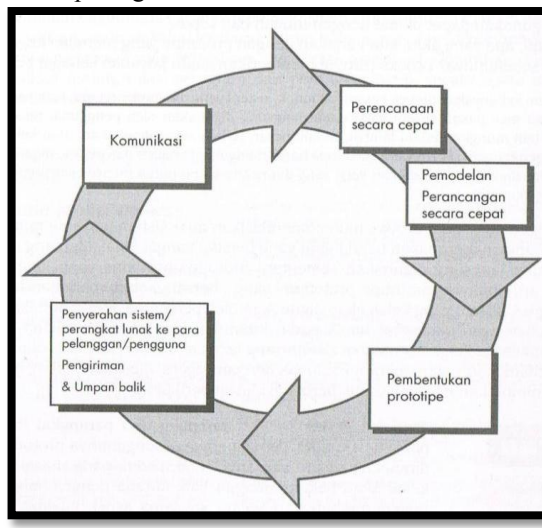

**Gambar 1.** Metode Pengembangan *Prototype*.

## 3 Hasil dan Pembahasan

Hasil dari penelitian yang telah dilakukan berupa sebuah aplikasi *e-service* berbasis *mobile Android* pada divisi pelayanan perbaikan komputer CV RKU. Aplikasi ini membantu pihak RKU dalam melayani pelanggan yang ingin melakukan perbaikan-perbaikan.

### 3.1 Halaman Login dan Home

Halaman login pada *Android* menampilkan menu *username* dan *password* pelanggan meng-*input*-kan data *username*-nya dengan mengetikkan nama dan email yang telah di data oleh *admin*.

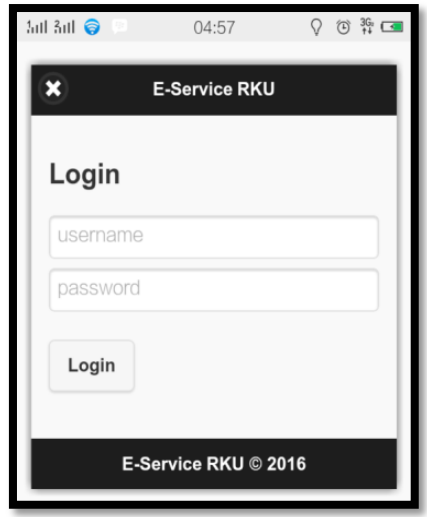

**Gambar 1.** Halaman Login dan Home.

### 3.2 Halaman *Service*

Halaman *service* merupakan halaman utama yang berfungsi bagi pelanggan untuk melihat informasi *service* barang yang sedang diperbaiki melalui *gadget Android*. Masukkan no nota, kemudian klik *button* "Cari", maka akan tampilkan data yang diinginkan. Halaman *service* nampak pada gambar 2.

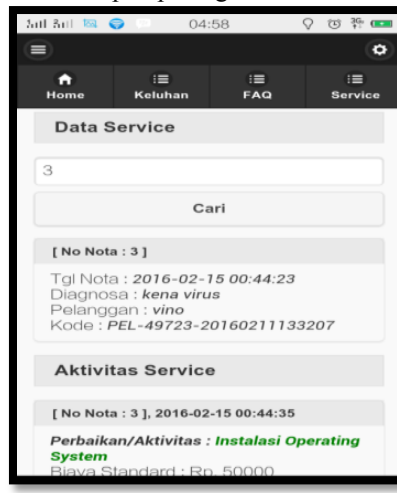

**Gambar 2.** Halaman *Service*.

### 3.3 Halaman Keluhan Pelanggan

Aplikasi *e-service* berbasis Android ini juga dilengkapi dengan fitur keluhan pelanggan. Halaman ini berfungsi bagi pelanggan untuk memasukkan keluhan-keluhannya. Dan bagi pihak RKU dapat melihat daftar keluhan pelanggan yang sudah di-*input*-kan oleh pelanggan.

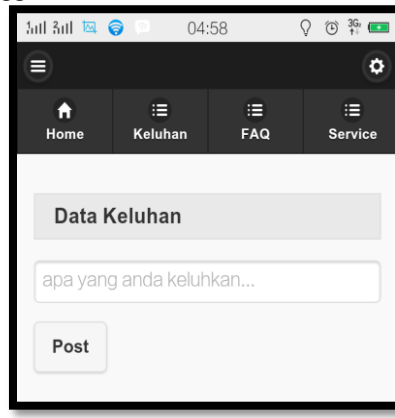

**Gambar 3.** Halaman Keluhan Pelanggan.

### 3.4 Halaman *Frequent Asked Questions (FAQ)*

Aplikasi *e-service* ini juga diperkaya denan fitur FAQ. Fitur ini berisi sejumlah pertanyaan yang sering diajukan oleh pelanggan. Sehingga pelanggan dapat melihat fitur ini sebagai pedoman untuk pertanyaan-pertanyaan yang umum diajukan oleh pelanggan-pelanggan sebelumnya.

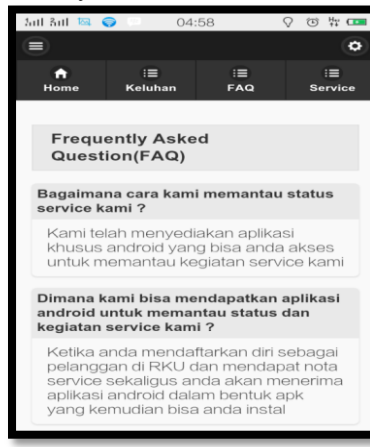

**Gambar 4.** Halaman *FAQ*.

## 4  Simpulan dan Saran

Berdasarkan hasil penelitian dan pembahasan mengenai penerapan sistem *e-service* pada divisi pelayanan perbaikan komputer CV RKU berbasis *Android*, dapat ditarik kesimpulan sebagai berikut :
1) *E-service* ini dirancang merupakan salah satu fasilitas tambahan dan nilai tambah bagi CV RKU untuk membantu mempermudah pelanggan dalam melaporkan keluhan yang ada di dalam komputer dan *printer*-nya dan membantu perusahaan dalam pelayanan pelanggan.
2) *E-service* ini dilengkapi dengan dengan berbasis *mobile* agar pelanggan dapat dengan mudah memperbaiki komputer dan *printer*-nya hanya dengan menggunakan layanan internet dan membuka *website* tersebut.

## Daftar Pustaka